\documentclass{iopjournal}

\usepackage{graphicx}
\usepackage{dcolumn}
\usepackage{bm}
\usepackage{hyperref}
\hypersetup{
    colorlinks=true,
    linkcolor=black,          
    urlcolor=blue,            
    citecolor=blue           
}
\usepackage[utf8]{inputenc}
\usepackage[T1]{fontenc}
\usepackage{mathptmx}
\usepackage{etoolbox}
\usepackage{amsmath}  
\usepackage{setspace}
\usepackage[numbers]{natbib}
\graphicspath{{figures/}}
\usepackage{float} 
\usepackage{graphicx}

\usepackage{caption,subcaption}
\begin{document}

\articletype{Paper}

\title{X-Ray Spectroscopy for Characterizing Stimulated Raman Scattering in Inertial Confinement Fusion}

\author{Tianluo Luo$^1$, Zeyang Li$^1$, Yunping Wang$^{1,2}$,
Zhihao Yang$^3$, Zhencen He$^1$, Dong Yang$^{2,*}$ and Zhimin Hu$^{1,*}$}

\affil{$^1$Key Laboratory of Radiation Physics and Technology of Ministry of Education, Institute of Nuclear Science and Technology, Sichuan University,
Chengdu 610064, China} \\
\affil{$^2$National Key Laboratory of Plasma Physics,
Laser Fusion Research Center, China Academy of Engineering Physics,
Mianyang, China} \\
\affil{$^3$College of Intelligent Manufacturing, Sichuan University of Arts and Science, Dazhou 635000, China}

\email{yangdong1@caep.cn; huzhimin@scu.edu.cn}


\begin{abstract}
Accurate characterization of stimulated Raman scattering (SRS) remains a critical challenge in inertial confinement fusion (ICF), as SRS not only scatters laser energy but also generates suprathermal electrons that preheat the fuel and degrade implosion performance. Conventional backscatter diagnostics provide direct measurements of SRS but cannot collect the entire scattered-light signal, limiting the accurate characterization of SRS strength. Using a non-local thermodynamic equilibrium collisional-radiative model with a double-Maxwellian electron distribution, we systematically investigate how suprathermal electrons modify the titanium \textit{K}-shell x-ray emission spectra. The spectra exhibit high sensitivity to the suprathermal-electron fraction at relatively low bulk electron temperatures, making them particularly suitable for diagnosing suprathermal electrons during the early stage of ICF, when even a small suprathermal-electron population can compromise fuel compression. The calculated spectra reproduce experimental measurements from the Nova Laser Facility with good agreement, and the inferred suprathermal-electron fractions show improved consistency with independently measured SRS losses compared with the original spectral analysis by Glenzer~\href{https://doi.org/10.1103/PhysRevLett.81.365} {\text{[S. H. Glenzer \textit{et al}., Phys. Rev. Lett. \textbf{81}, 365(1998)]}}. These results demonstrate that \textit{K}-shell spectroscopy, combined with accurate NLTE collisional-radiative modeling, provides a reliable probe of SRS strength in laser-produced plasmas. Hence, this approach may be extended to spatially resolved diagnosis of SRS strength, offering a promising complement to conventional backscatter diagnostics.
\end{abstract}

\section{INTRODUCTION}

With the recent achievement of ignition and burning-plasma conditions at the National Ignition Facility, accurate diagnosis of high-energy-density plasmas has become increasingly important\cite{PhysRevLett.132.065102,PhysRevLett.129.075001,Zylstra2022}. Laser--plasma instabilities remain one of the central issues in inertial confinement fusion (ICF). In particular, stimulated Raman scattering (SRS) and two-plasmon decay can produce suprathermal electrons and give rise to non-Maxwellian electron energy distributions (EEDFs)~\cite{PhysRevLett.49.371,PhysRevA.6.764,PhysRevLett.108.175002}. These energetic electrons can preheat the fuel and reduce its compressibility, thereby degrading implosion performance\cite{Key91,Lang80,PhysRevLett.120.055001,RevModPhys.95.025005,10.1063/5.0152191}. Conventional backscatter measurements provide information on the scattered light escaping from the target, but do not directly probe the suprathermal-electron fraction in the plasma. Diagnostics sensitive to the suprathermal electrons themselves can therefore provide complementary information on the effects of laser--plasma instabilities within the plasma. Moreover, by modifying the EEDF, suprathermal electrons alter electron-driven processes and consequently affect charge-state distribution (CSD) and radiation properties in plasmas~\cite{HS04}.

The sensitivity of atomic kinetics to the EEDF has been demonstrated in a wide range of laboratory plasmas. Garland \textit{et al}. demonstrated that minority relativistic electron populations can strongly affect the charge-state balance and radiation losses in post-disruption tokamak plasmas~\cite{Garland22}. Bao \textit{et al}. further showed that suprathermal electrons remarkably enhance the electron collisional ionization rate coefficients of highly charged tungsten ions at relatively low bulk electron temperatures~\cite{Bao23}. Experimental and numerical studies of aluminum, argon, magnesium, and beryllium plasmas have also shown that suprathermal electrons can substantially modify the relative intensities of \textit{K}-shell satellite lines through their effects on collisional excitation and ionization processes~\cite{JAb96,Abd99,JAb99}. These studies demonstrate the importance of accurately accounting for the EEDF in collisional-radiative (CR) modeling and spectroscopic plasma diagnostics.

The use of \textit{K}-shell spectroscopy for suprathermal-electron diagnostics was established in earlier studies. Rosmej showed that a two-temperature EEDF can simultaneously enhance H-like emission while retaining satellite emission from lower charge states, providing a characteristic spectroscopic signature of a suprathermal-electron population~\cite{Rosmej1997}. Glenzer \textit{et al}. subsequently applied a time-dependent non-Maxwellian CR model to titanium \textit{K}-shell spectra measured in Nova hohlraum plasmas and inferred the temporal evolution of the suprathermal-electron fraction~\cite{SH98}. These studies established the physical basis of \textit{K}-shell suprathermal-electron spectroscopy. However, the sensitivity of $K$-shell spectra to suprathermal electrons strongly depends on the bulk electron temperature. A systematic understanding of this temperature dependence is therefore important for the spectroscopic determination of the suprathermal-electron fraction. In addition, it remains to be explored whether \textit{K}-shell spectroscopy of suprathermal electrons can complement conventional SRS backscatter measurements and thereby provide a more comprehensive characterization of SRS in hohlraum plasmas.

In this work, we employ a non-local thermodynamic equilibrium (NLTE) collisional-radiative model with a double-Maxwellian EEDF to systematically investigate the effects of suprathermal electrons on titanium \textit{K}-shell x-ray emission under ICF-relevant conditions. We examine how the suprathermal-electron fraction and bulk electron temperature affect atomic rate coefficients, charge-state distributions, and the resulting spectral features, with particular emphasis on the plasma conditions under which the titanium spectrum retains sensitivity to the suprathermal-electron population. The calculated spectra are then compared with experimental measurements from Nova hohlraum plasmas to infer the temporal evolution of the suprathermal-electron fraction. Finally, the inferred suprathermal-electron fractions are compared with independently measured SRS losses to assess the potential of \textit{K}-shell spectroscopy as a complementary diagnostic of SRS-associated suprathermal electrons in ICF plasmas.

\section{THEORY}
\subsection{Non-Maxwellian electron distribution}
For a Maxwellian population with temperature $T$ expressed in energy units, the normalized electron energy distribution is
\begin{equation}
F(E; T) = 2 \sqrt{\frac{E}{\pi T^3}} \exp\left(-\frac{E}{T}\right),
\end{equation}
where $E=m_ev^2/2$ is the electron kinetic energy. We represent the EEDF in the emitting region by a bulk Maxwellian and a minority suprathermal Maxwellian,
\begin{equation}
F_{\text{tot}}(E) = (1-f)F(E;T_e)+fF(E;T_{\rm hot}),
\end{equation}
where $T_e$ and $T_{\rm hot}$ are the bulk and suprathermal-electron temperatures, respectively. The local suprathermal-electron number fraction is defined as~\cite{SH98}
\begin{equation}
f=\frac{n_{\rm hot}}{n_{\rm bulk}+n_{\rm hot}}.
\end{equation}
The double-Maxwellian distribution is a compact parameterization of the energetic tail rather than a kinetic model of its generation. It does not describe angular anisotropy or polarization. Following the Nova analysis, we fix $T_{\rm hot}=19$ keV and examine how the spectral response varies with $T_e$ and $f$.

\subsection{Atomic data and level populations}
Atomic structure and cross sections of different atomic processes are calculated with the Flexible Atomic Code (FAC)~\cite{GM08} using a relativistic configuration-interaction method. An atomic state function $\Psi$ with parity $\Pi$, total angular momentum $J$, magnetic quantum number $M$, and additional quantum number $\gamma$ is expanded in configuration state functions (CSFs) $\Phi$ with the same $\Pi JM$:
\begin{equation}
\Psi(\gamma\Pi J M)=\sum_{r=1}^{n} c_{r} \Phi\left(\gamma_{r} \Pi J M\right),
\end{equation}
where $c_r$ is obtained by diagonalizing the model Hamiltonian. The configuration expansion includes titanium ions from H-like through C-like. For each ionization stage, we include the ground configuration, singly excited configurations through principal quantum number $n=6$, $K$-shell-vacancy configurations, and the associated autoionizing configurations.

The Hamiltonian (in atomic units) for an atomic ion with $N$ electrons can be written as:
\begin{equation}
H=\sum_{i=1}^{N} H_{D}(i)+\sum_{i<j} V_{i j}^{e e},
\end{equation}
where $H_D(i)$ is the Dirac Hamiltonian for electron $i$, and $V_{ij}^{ee}$ is the electron--electron interaction, including the Breit interaction~\cite{BG29}.

In an NLTE plasma, the ionic level populations obey the collisional-radiative rate equation
\begin{equation}
\frac{\mathrm{d}n_{i}}{\mathrm{d}t}=\sum_{j}R_{ji}n_{j}-\sum_{j}R_{ij}n_{i},
\label{eq:CR}
\end{equation}
where $R_{ij}$ is the rate from level $i$ to level $j$. The model includes electron collisional excitation (CE) and de-excitation, collisional ionization (CI) and three-body recombination, dielectronic capture (DC) and autoionization, radiative recombination (RR) and photoionization, and photoexcitation and radiative decay.

The non-Maxwellian EEDF is coupled to the CR model through the electron-driven rate coefficients. For an atomic process $p$ involving
a single incident electron, the rate coefficient for a transition from
state $i$ to state $j$ is obtained by averaging the corresponding cross
section over the EEDF,
\begin{equation}
K_{ij}^{(p)}[F_{\rm tot}]
=
\int_{E_{\rm th}^{(p)}}^{\infty}
\sigma_{ij}^{(p)}(E)\,
v(E)\,
F_{\rm tot}(E)\,
\mathrm{d}E ,
\label{eq:general_rate}
\end{equation}
where $p$ labels the relevant electron-driven process, $\sigma_{ij}^{(p)}(E)$ is the corresponding cross section,
$v(E)=\sqrt{2E/m_e}$ is the incident-electron speed, and
$E_{\rm th}^{(p)}$ is the threshold energy. Because the distribution
in Eq.~(2) is a linear combination of two normalized Maxwellian
distributions, the resulting rate coefficient can equivalently be
written as
\begin{equation}
K_{ij}^{(p)}
=
(1-f)K_{ij}^{(p)}(T_e)
+
fK_{ij}^{(p)}(T_{\rm hot}).
\label{eq:double_maxwellian_rate}
\end{equation}
Autoionization rates and Einstein $A$ coefficients are obtained
directly from the atomic-structure calculation. Three-body
recombination, which involves two free electrons, is treated
separately using the prescription implemented in the CR model. Further details of the atomic-rate calculations and their numerical implementation are given
in our previous work~\cite{Luo25}.

By solving the above equations, the population numbers $n_i$ can be obtained. This allows us to determine the charge-state distribution $F$ of ions with nuclear charge $Z$:
\begin{equation}
F(N) = \frac{\sum_i n(N,i)}{\sum_N \sum_i n(N,i)}.
\end{equation}

\subsection{Spectral synthesis}
The frequency-dependent volume absorption coefficient $\kappa_\nu$ and emission coefficient $j_\nu$ contain free--free (ff), bound--free (bf), and bound--bound (bb) contributions,
\begin{equation}
\begin{aligned}
\kappa_\nu&=\kappa_\nu^{ff}+\kappa_\nu^{bf}+\kappa_\nu^{bb},\\
j_\nu&=j_\nu^{ff}+j_\nu^{bf}+j_\nu^{bb}.
\end{aligned}
\label{eq:total_coefficients}
\end{equation}
Since the present work focuses on titanium $K$-shell line emission, only the bound--bound contributions are given explicitly below. The free--free and bound--free continua are calculated using the standard formulations adopted in our previous work~\cite{Luo25}.

For a bound--bound transition from an upper level $i$ to a lower level $j$, the absorption coefficient is expressed in terms of the spontaneous radiative-decay rate as
\begin{equation}
\kappa_\nu^{bb}
=\sum_N\sum_{i,j}
n(N,j)
\frac{g_i}{g_j}
\frac{c^2}{8\pi\nu_{ij}^{2}}
A_{ij}\phi_{ij}(\nu),
\label{eq:bb_absorption}
\end{equation}
and the spontaneous emission coefficient is
\begin{equation}
j_\nu^{bb}
=\frac{1}{4\pi}
\sum_N\sum_{i,j}
n(N,i)A_{ij}h\nu_{ij}\phi_{ij}(\nu),
\label{eq:bb_emission}
\end{equation}
where $A_{ij}$ is the spontaneous radiative-decay rate from level $i$ to level $j$, $g_i$ and $g_j$ are the statistical weights of the upper and lower levels, respectively, and $\phi_{ij}(\nu)$ is normalized such that $\int\phi_{ij}(\nu)\,\mathrm{d}\nu=1$. For a uniform emitting region with an effective photon path length $L_{\rm eff}$ and negligible incident background, the emergent specific intensity is
\begin{equation}
I_\nu=S_\nu\left(1-e^{-\tau_\nu}\right),
\qquad
S_\nu=\frac{j_\nu}{\kappa_\nu},
\qquad
\tau_\nu=\kappa_\nu L_{\rm eff}.
\label{eq:transfer}
\end{equation}

To compare with the experimental spectra, each transition is broadened by a Voigt profile,
\begin{equation}
\begin{aligned}
\phi_{ij}(\nu)
&=
\frac{2\ln2}{\pi\sqrt{\pi}}
\frac{\Gamma_{\nu,ij}}{\omega_\nu^{2}}
\int_{-\infty}^{\infty}
\frac{\exp(-t^{2})\,\mathrm{d}t}
{\left[2\sqrt{\ln2}(\nu-\nu_{ij})/\omega_\nu-t\right]^{2}
+\left(\sqrt{\ln2}\Gamma_{\nu,ij}/\omega_\nu\right)^{2}},
\end{aligned}
\label{eq:voigt}
\end{equation}
where $\Gamma_{\nu,ij}$ is the Lorentzian natural linewidth and
$\omega_\nu$ is the Gaussian full width at half maximum, including
Doppler and instrumental broadening. If the widths are specified in energy units, the corresponding frequency widths are obtained by division by $h$.

\section{RESULTS AND DISCUSSION}

\subsection{Response of the atomic rate coefficients}

Unless stated otherwise, the calculations use an electron density of $n_{\rm e}=1\times10^{21}\ \mathrm{cm}^{-3}$ and a suprathermal-electron temperature of $T_{\rm hot}=19$ keV. We first examine the elementary rate coefficients in order to identify how the EEDF modifies the charge-state balance and, ultimately, the spectrum.

\begin{figure}
 \centering
        \includegraphics[width=1\textwidth]{figure2.PNG}
 \caption{Representative electron collisional excitation (top) and collisional-ionization (bottom) rate coefficients for H-like, He-like, and Li-like titanium calculated with the double-Maxwellian EEDF.}
\label{fig:2}
\end{figure}

Figure~\ref{fig:2} shows the electron CE and CI rate coefficients of H-like, He-like, and Li-like titanium ions calculated with the double-Maxwellian electron energy distribution for different bulk electron temperatures and suprathermal-electron fractions. As the number of bound electrons increases from H-like to Li-like ions, the overall CE and CI rate coefficients become larger because more electron collisional excitation and ionization channels are available. For all three charge states, the CE rate coefficients decrease with increasing bulk electron temperature. At a given bulk electron temperature, they are further reduced as the suprathermal electron fraction increases. The CI rate coefficients of He-like and Li-like ions exhibit the same behavior. In contrast, the CI rate coefficient of H-like ions increases with both the bulk electron temperature and the suprathermal-electron fraction. More importantly, the effect of the suprathermal electron fraction on all these rate coefficients gradually weakens as the bulk electron temperature increases.

The dielectronic-capture coefficients are approximately three orders of magnitude smaller than the displayed CE and CI coefficients under these conditions and are therefore not shown. Autoionization and photoionization do not involve an incident free electron and are not directly modified by the assumed EEDF. The rate trends in Fig.~\ref{fig:2} thus anticipate that the influence of the suprathermal component should be most apparent at the lower bulk temperatures.

\subsection{Temperature-dependent spectral sensitivity and ionization balance}

\begin{figure}[t]
 \centering
        \includegraphics[width=1.0\textwidth]{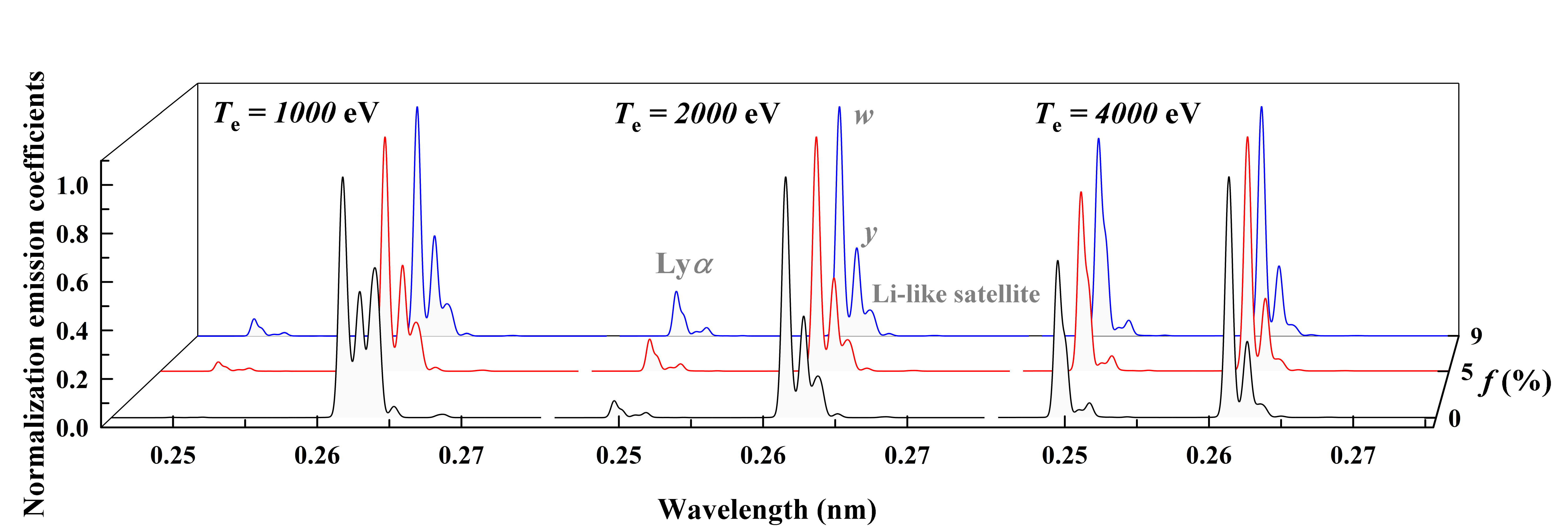}
 \caption{$K$-shell X-ray emission coefficients of titanium plasma with different bulk electron temperatures and suprathermal-electron fractions.}
\label{fig:3}
\end{figure}

Figure~\ref{fig:3} presents the calculated \textit{K}-shell emission coefficients for suprathermal-electron fractions of $f=0\%$, $f=5\%$, and $f=9\%$ at bulk electron temperatures from 1 to 4 keV. The selected wavelength range of 0.245--0.275 nm covers the H-like Ly-$\alpha$, He-like He-$\alpha$, and Li-like satellite features. The satellites weaken and Ly-$\alpha$ strengthens as $T_e$ rises, reflecting the shift toward higher average charge. At $T_e=1$ keV, increasing $f$ produces pronounced changes in both features. These differences decrease at 2 keV and are comparatively small at 4 keV, consistent with the convergence of the rate coefficients in Fig.~\ref{fig:2}. The titanium spectrum is therefore substantially more sensitive to the suprathermal fraction at the lower bulk temperatures considered here. Titanium \textit{K}-shell spectroscopy is most responsive to a suprathermal component when the bulk population alone does not already sustain the highest charge states efficiently. This high sensitivity in the low-temperature regime is particularly relevant to ICF, where suprathermal-electron preheating of the relatively cold fuel during the early stage of the implosion can increase the fuel adiabat and reduce capsule compressibility. Titanium \textit{K}-shell spectroscopy is therefore particularly suitable for diagnosing suprathermal electrons during this critical stage.

\begin{figure}[t]
 \centering
        \includegraphics[width=0.6\textwidth]{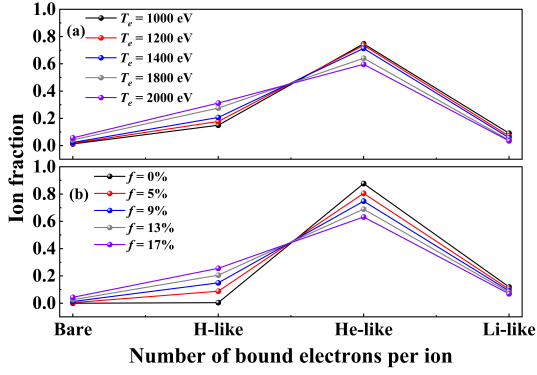}
 \caption{CSDs of titanium ions for (a) different bulk electron temperatures at a fixed suprathermal-electron fraction of f=$9\%$, and (b) different suprathermal-electron fractions at a fixed bulk electron temperature of $T_{\rm e}=1000$ eV.}
\label{fig:4}
\end{figure}

To further understand how suprathermal electrons modify the emission spectra of titanium plasma, we calculated the corresponding ionic charge-state distributions. Figure~\ref{fig:4} presents the charge-state distributions corresponding to the spectra shown in Fig.~\ref{fig:3}, where the horizontal axis denotes the number of bound electrons of the titanium ions. Figure~\ref{fig:4}(a) shows the charge-state distributions calculated with a fixed suprathermal-electron fraction of $f=9\%$ at different bulk electron temperatures. As the bulk electron temperature increases, the ionization balance shifts toward higher charge states, with increasing bare- and H-like ion fractions and decreasing He-like and Li-like fractions. Consequently, the enhanced H-like population strengthens the Ly-$\alpha$ emission, while the reduced He-like and Li-like populations weaken the corresponding He-$\alpha$, intercombination, and satellite features, consistent with the spectral evolution shown in Fig.~\ref{fig:3}.

Figure~\ref{fig:4}(b) shows the charge-state distributions for different suprathermal-electron fractions at a fixed bulk electron temperature of $T_{\rm e}=1000$ eV. Increasing the suprathermal-electron fraction produces the same overall trend as increasing the bulk electron temperature, namely a redistribution toward higher charge states. This similarity explains why both parameters strengthen Ly-$\alpha$ emission and weaken Li-like satellites in Fig.~\ref{fig:3}. It also implies a partial degeneracy between $T_e$ and $f$, which must be considered when spectra are used for parameter inference. Such a degeneracy can, however, be partially resolved by simultaneously considering spectral features from a broader range of charge states. As pointed out by Rosmej~\cite{Rosmej1997}, the presence of a suprathermal-electron component can enhance emission from highly charged ions while substantial populations of lower charge states are still retained by the relatively cool bulk electrons. In contrast, increasing the bulk electron temperature shifts the overall ionization balance toward higher charge states and therefore strongly suppresses the emission from lower ionization stages. The simultaneous observation of highly ionized species and lower-charge-state $K$-shell satellites can thus help distinguish the effects of the suprathermal-electron fraction from those of the bulk electron temperature.

\subsection{Reanalysis of the Nova spectra}

\begin{figure}[!t]
 \centering
        \includegraphics[width=0.72\textwidth]{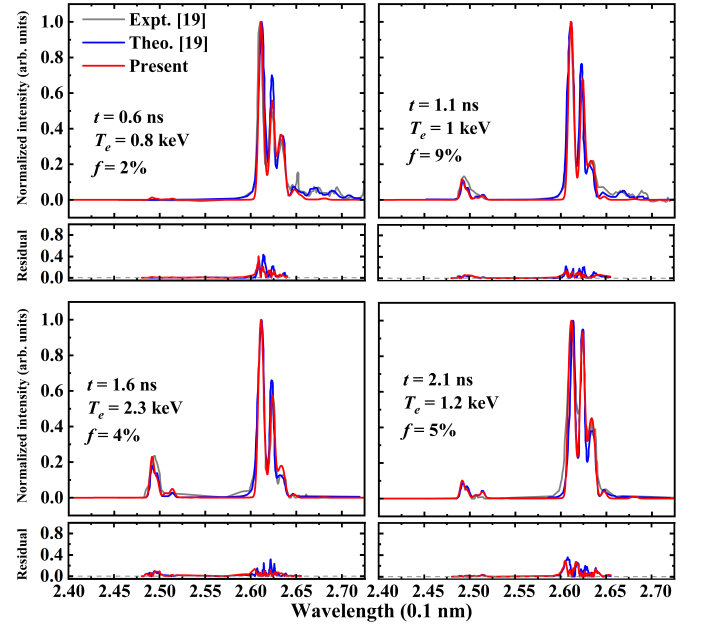}
 \caption{$K$-shell x-ray spectra of titanium plasmas at different times. The gray and blue lines represent the experimental and theoretical spectra reported in Ref.~\cite{SH98}, respectively, while the red lines denote the present calculations. The lower panels show the corresponding absolute residuals, $\Delta I=|I_{\rm cal}-I_{\rm exp}|$, with the dashed lines indicating $\Delta I=0$.}
\label{fig:5}
\end{figure}

Based on the above analysis of the effects of suprathermal-electron fraction and bulk electron temperature on the titanium \textit{K}-shell spectra, we recalculated the emission spectra under the experimental conditions of Glenzer \textit{et al}.~\cite{SH98}. The calculated spectra are compared with the experimental measurements at 0.6, 1.1, 1.6, and 2.1 ns, as shown in Fig.~\ref{fig:5}. The gray and blue lines represent the experimental spectra and the previous theoretical results obtained from Ref.~\cite{SH98}, respectively, while the red solid lines denote the present calculations. Opacity effects were included through an effective photon path length, with $L_{\rm eff}=200~\mu$m adopted for the first three time instants. Compared with the emission coefficients shown in Fig.~\ref{fig:3}, opacity mainly modifies the relative intensities of the He-like resonance line ($w$), the intercombination line ($y$), and the Li-like satellite lines. In particular, the larger absorption coefficient of the $w$ line results in stronger attenuation of this transition relative to the other spectral features.

The present double-Maxwellian NLTE collisional-radiative model reproduces the measured spectra well over the entire temporal evolution. The absolute residuals, $\Delta I=|I_{\rm cal}-I_{\rm exp}|$, shown in the lower panels of Fig.~\ref{fig:5}, are generally smaller than those of the previous calculations. This comparison demonstrates the improved spectral agreement achieved by the present NLTE collisional-radiative modeling.

\begin{figure}[!t]
 \centering
        \includegraphics[width=0.58\textwidth]{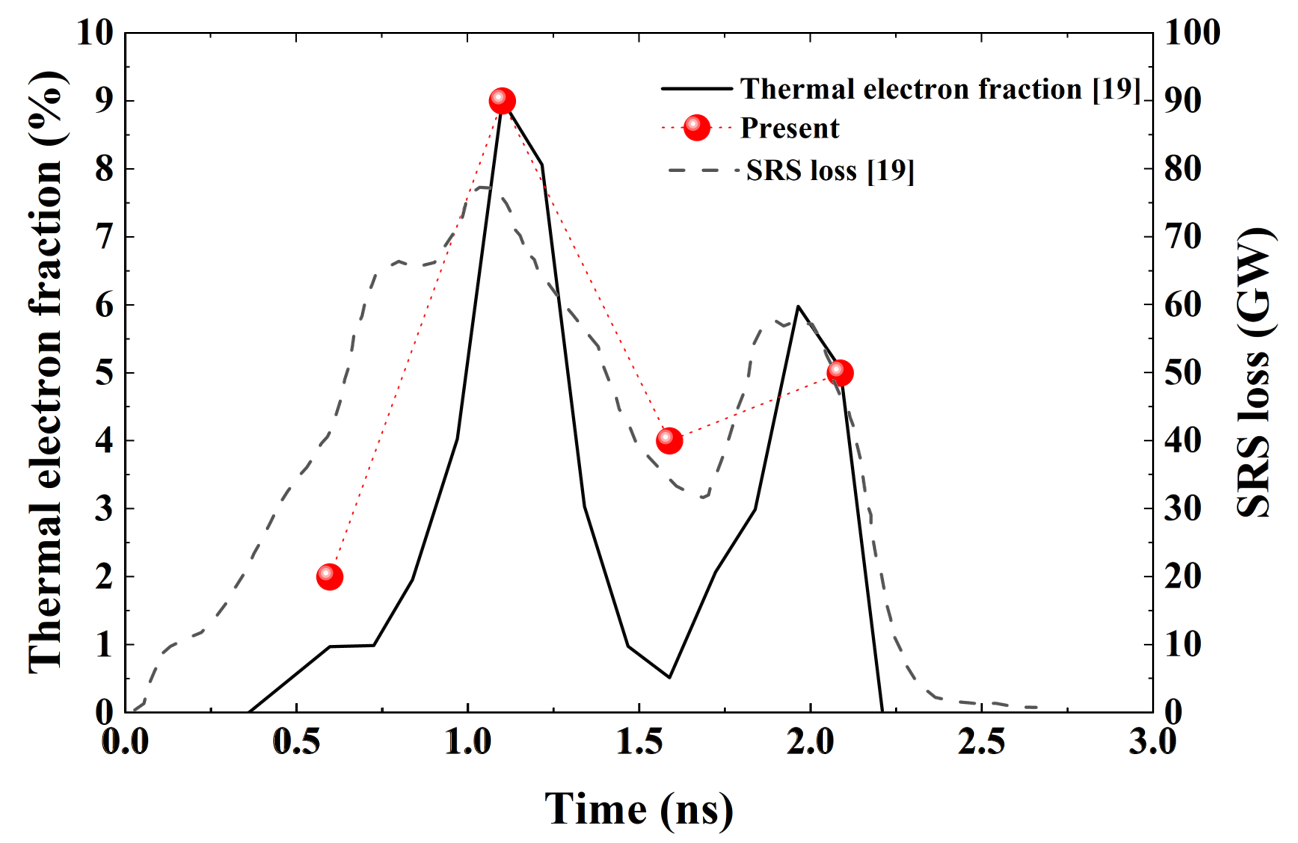}
 \caption{Temporal evolution of the suprathermal-electron fraction and its correlation with the SRS losses reported in Ref.~\cite{SH98}.}
\label{fig:6}
\end{figure}

The improved spectral modeling also allows the suprathermal-electron fraction to be more accurately determined. Figure~\ref{fig:6} compares its temporal evolution obtained from the present spectral fitting with that reported in Ref.~\cite{SH98} and the independently measured SRS losses. The suprathermal-electron fractions obtained from the present model show a qualitatively similar trend as the measured SRS losses, with both quantities reaching their maximum values near 1.1 ns and decreasing afterwards. Compared with the previous results, the present results show better consistency with the measured SRS evolution, particularly at 1.6 and 2.1 ns. This agreement provides an independent validation of the suprathermal-electron fractions obtained from the present spectral analysis. Together with the improved agreement between the calculated and measured spectra in Fig.~\ref{fig:5}, these results support the reliability of the present NLTE collisional-radiative model for quantitatively diagnosing the local suprathermal-electron population.

The present results also suggest that \textit{K}-shell spectroscopy can be extended to the spatially resolved diagnosis of laser--plasma instabilities. Previous experiments have demonstrated that spatially separated mid-$Z$ spectroscopic tracer targets can provide spatially resolved measurements inside ignition hohlraums~\cite{PhysRevLett.121.095002}. Since titanium $K$-shell spectra are sensitive to the local suprathermal-electron fraction, a similar platform could be used to infer the spatial distribution of suprathermal electrons and thereby provide spatially resolved information on SRS strength that is difficult to obtain with conventional backscatter diagnostics. Furthermore, the bulk electron temperature can vary considerably across different regions of an ignition hohlraum, and the tracer element should therefore be selected according to the local plasma temperature. For the titanium considered here, the \textit{K}-shell spectrum is most sensitive to the suprathermal-electron fraction at bulk electron temperature approximately 1000--2000 eV, while higher-$Z$ elements may be required above 2000 eV to maintain sufficient sensitivity to the suprathermal-electron fraction.

\section{CONCLUSION}
We have demonstrated that titanium \textit{K}-shell x-ray spectroscopy provides a powerful tool for characterizing SRS-generated suprathermal electrons in ICF plasmas. Using a double-Maxwellian NLTE collisional-radiative model, we systematically investigated how suprathermal electrons modify the \textit{K}-shell emission spectra. The spectra are highly sensitive to the suprathermal-electron fraction at low bulk electron temperatures, making \textit{K}-shell spectroscopy particularly suitable for diagnosing the early stage of ICF implosions, where fuel preheat strongly affects capsule compression. The calculated spectra reproduce the Nova experimental measurements with good agreement, and the inferred suprathermal-electron fractions show improved consistency with independently measured SRS losses compared with the original spectral analysis. These results demonstrate that accurate NLTE collisional-radiative modeling combined with \textit{K}-shell spectroscopy provides a reliable probe of SRS strength. Furthermore, spatially resolved spectroscopic tracer targets may extend this approach toward spatially resolved diagnosis of SRS in laser-produced plasmas.

\funding{This work was supported by the Science Challenge Project (Grant No. TZ2025014), the Fund of National Key Laboratory of Plasma Physics (Grant No. 6142A042420101), the National Natural Science Foundation of China (Grant No. 12304276), and the Sichuan Science and Technology Program (Grant Nos. 2026NSFSC0811 and 2025NSFJQ0042).}

\data{The data that support the findings of this study are available from the corresponding author upon reasonable request.}

  \bibliographystyle{unsrt}
\bibliography{reference}

@PREAMBLE{
 "\providecommand{\noopsort}[1]{}" 
 # "\providecommand{\singleletter}[1]{#1}%" 
}

@article{Rosmej1997,
  author  = {Rosmej, F. B.},
  title   = {Hot electron x-ray diagnostics},
  journal = {J. Phys. B: At. Mol. Opt. Phys.},
  volume  = {30},
  pages   = {L819--L828},
  year    = {1997},
  doi     = {10.1088/0953-4075/30/22/007}
}

@article{Key91, 
title={The Physics of Laser Plasma Interactions, W. L. Kruer. Addison-Wesley, 1988, £33.95, 182 pages.}, 
volume={45},
journal={Journal of Plasma Physics}, 
author={Key, M. H.}, 
year={1991}, 
pages={135--135},
DOI={10.1017/S0022377800015555},
}

@article{Lang80,
  title = {Nonlinear Inverse Bremsstrahlung and Heated-Electron Distributions},
  author = {Langdon, A. Bruce},
  journal = {Phys. Rev. Lett.},
  volume = {44},
  pages = {575--579},
  year = {1980},
    doi = {10.1103/PhysRevLett.44.575},
  url = {https://link.aps.org/doi/10.1103/PhysRevLett.44.575}
}

@article{HS04,
title = {Effects of the electron energy distribution function on modeled x-ray spectra},
  author = {Hansen, S. B. and Shlyaptseva, A. S.},
  journal = {Phys. Rev. E},
  volume = {70},
  issue = {3},
  pages = {036402},
  numpages = {13},
  year = {2004},
  month = {Sep},
  publisher = {American Physical Society},
  doi = {10.1103/PhysRevE.70.036402},
  url = {https://link.aps.org/doi/10.1103/PhysRevE.70.036402}
}

@ARTICLE{SH98,
author = "S. H. Glenzer and F. B. Rosmej and R. W. Lee and C. A. Back and K. G. Estabrook and B. J. MacGowan and T. D. Shepard and R. E. Turner",
year = "1998",
journal = "Phys.\ Rev.\ Lett.",
volume = "81",
title="Measurements of Suprathermal Electrons in Hohlraum Plasmas with X-Ray Spectroscopy. ",
pages = "365",
publisher = {American Physical Society},
doi = {10.1103/PhysRevLett.81.365},
url = {https://link.aps.org/doi/10.1103/PhysRevLett.81.365}
}

@article{GM08,
author = {Gu, M F},

title = {The flexible atomic code},
journal = {Canadian Journal of Physics},
volume = {86},
number = {5},
pages = {675-689},
year = {2008},
doi = {10.1139/p07-197},
URL = {https://doi.org/10.1139/p07-197},
}

@article{BG29,
  title = {The Effect of Retardation on the Interaction of Two Electrons},
  author = {Breit, G.},
  journal = {Phys. Rev.},
  volume = {34},
  pages = {553--573},
  year = {1929},
    doi = {10.1103/PhysRev.34.553},
  url = {https://link.aps.org/doi/10.1103/PhysRev.34.553}
}

@article{Luo25,
  title = {Radiation field effect on the Fe plasma spectra in the Perseus cluster core},
  author = {Luo, Tianluo and He, Zhencen and Liu, Xianliang and Yang, Zhihao and Zhang, Shuyu and Hu, Zhimin},
  journal = {Phys. Rev. E},
  volume = {112},
  pages = {065203},
  year = {2025},
  publisher = {American Physical Society},
  doi = {10.1103/k59l-7yml},
  url = {https://link.aps.org/doi/10.1103/k59l-7yml}
}

@article{PhysRevLett.49.371,
  title = {Stimulated Raman Backscatter from a Magnetically Confined Plasma Column},
  author = {Offenberger, A. A. and Fedosejevs, R. and Tighe, W. and Rozmus, W.},
  journal = {Phys. Rev. Lett.},
  volume = {49},
  issue = {6},
  pages = {371--375},
  numpages = {0},
  year = {1982},
  month = {Aug},
  publisher = {American Physical Society},
  doi = {10.1103/PhysRevLett.49.371},
  url = {https://link.aps.org/doi/10.1103/PhysRevLett.49.371}
}

@article{PhysRevA.6.764,
  title = {Experimental Indications of Plasma Instabilities Induced by Laser Heating},
  author = {Shearer, J. W. and Mead, S. W. and Petruzzi, J. and Rainer, F. and Swain, J. E. and Violet, C. E.},
  journal = {Phys. Rev. A},
  volume = {6},
  issue = {2},
  pages = {764--769},
  numpages = {0},
  year = {1972},
  month = {Aug},
  publisher = {American Physical Society},
  doi = {10.1103/PhysRevA.6.764},
  url = {https://link.aps.org/doi/10.1103/PhysRevA.6.764}
}

@article{Bao23,
doi = {10.1088/1361-6587/acf15d},
url = {https://doi.org/10.1088/1361-6587/acf15d},
year = {2023},
month = {aug},
publisher = {IOP Publishing},
volume = {65},
number = {10},
pages = {105004},
author = {Bao, Runjia and Li, Bowen},
title = {Maxwellian and non-Maxwellian rate coefficients of tungsten ions from W46+ to W55+},
journal = {Plasma Physics and Controlled Fusion},
}

@article{Garland22,
  author  = {Garland, N. A. and Scott, H. A. and Hansen, S. B. and others},
  title   = {The impact of relativistic electrons on post-disruption tokamak plasmas},
  journal = {Physics of Plasmas},
  volume  = {29},
  pages   = {012504},
  year    = {2022},
  doi     = {10.1063/5.0076168}
}

@article{Abd99,
  author  = {Abdallah Jr, J. and Clark, R. E. H. and Faenov, A. Ya. and Karpinski, L. and Pikuz, S. A. and Romanova, V. M. and Sadowski, M. and Scholz, M. and Szydlowski, A.},
  title   = {Electron beam effects on the spectroscopy of multiply charged ions in plasma focus experiments},
  journal = {Journal of Quantitative Spectroscopy and Radiative Transfer},
  volume  = {62},
  pages   = {85--96},
  year    = {1999},
  doi     = {10.1016/S0022-4073(98)00073-9}
}

@article{PhysRevLett.108.175002,
  title = {Generating energetic electrons through staged acceleration in the two-plasmon-decay instability in inertial confinement fusion},
  author = {Yan, R. and Ren, C. and Li, J. and Maximov, A. V. and Mori, W. B. and Sheng, Z.-M. and Tsung, F. S.},
  journal = {Phys. Rev. Lett.},
  volume = {108},
  issue = {17},
  pages = {175002},
  numpages = {5},
  year = {2012},
  month = {Apr},
  publisher = {American Physical Society},
  doi = {10.1103/PhysRevLett.108.175002},
  url = {https://link.aps.org/doi/10.1103/PhysRevLett.108.175002}
}

@article{PhysRevLett.132.065102,
  title = {Achievement of Target Gain Larger than Unity in an Inertial Fusion Experiment},
  author = {Abu-Shawareb, H.},
  collaboration = {The Indirect Drive ICF Collaboration},
  journal = {Phys. Rev. Lett.},
  volume = {132},
  issue = {6},
  pages = {065102},
  numpages = {16},
  year = {2024},
  month = {Feb},
  publisher = {American Physical Society},
  doi = {10.1103/PhysRevLett.132.065102},
  url = {https://link.aps.org/doi/10.1103/PhysRevLett.132.065102}
}

@article{PhysRevLett.129.075001,
  title = {Lawson Criterion for Ignition Exceeded in an Inertial Fusion Experiment},
  author = {Abu-Shawareb, H. and Acree, R. and Adams, P.},
  collaboration = {Indirect Drive ICF Collaboration},
  journal = {Phys. Rev. Lett.},
  volume = {129},
  issue = {7},
  pages = {075001},
  numpages = {19},
  year = {2022},
  month = {Aug},
  publisher = {American Physical Society},
  doi = {10.1103/PhysRevLett.129.075001},
  url = {https://link.aps.org/doi/10.1103/PhysRevLett.129.075001}
}

@article{Zylstra2022,
author={Zylstra, A. B.
and Hurricane, O. A.
and Callahan, D. A.
and Kritcher, A. L.},
title={Burning plasma achieved in inertial fusion},
journal={Nature},
year={2022},
month={Jan},
day={01},
volume={601},
number={7894},
pages={542-548},
issn={1476-4687},
doi={10.1038/s41586-021-04281-w},
url={https://doi.org/10.1038/s41586-021-04281-w}
}

@article{RevModPhys.95.025005,
  title = {Physics principles of inertial confinement fusion and U.S. program overview},
  author = {Hurricane, O. A. and Patel, P. K. and Betti, R. and Froula, D. H. and Regan, S. P. and Slutz, S. A. and Gomez, M. R. and Sweeney, M. A.},
  journal = {Rev. Mod. Phys.},
  volume = {95},
  issue = {2},
  pages = {025005},
  numpages = {66},
  year = {2023},
  month = {Jun},
  publisher = {American Physical Society},
  doi = {10.1103/RevModPhys.95.025005},
  url = {https://link.aps.org/doi/10.1103/RevModPhys.95.025005}
}

@article{10.1063/5.0152191,
    author = {Rosenberg, M. J. and Solodov, A. A. and Stoeckl, C. and Hohenberger, M. and Bahukutumbi, R. and Theobald, W. and Edgell, D. and Filkins, T. and Betti, R. and Marshall, F. J. and Shah, R. C. and Turnbull, D. P. and Christopherson, A. R. and Lemos, N. and Tubman, E. and Regan, S. P.},
    title = {Hot electron preheat in hydrodynamically scaled direct-drive inertial confinement fusion implosions on the NIF and OMEGA},
    journal = {Physics of Plasmas},
    volume = {30},
    number = {7},
    pages = {072710},
    year = {2023},
    month = {07},
    issn = {1070-664X},
    doi = {10.1063/5.0152191},
    url = {https://doi.org/10.1063/5.0152191},
}

@article{PhysRevLett.121.095002,
  title = {Developing an Experimental Basis for Understanding Transport in NIF Hohlraum Plasmas},
  author = {Barrios, M. A. and Moody, J. D. and Suter, L. J. and Sherlock, M. and Chen, H. and Farmer, W. and Jaquez, J. and Jones, O. and Kauffman, R. L. and Kilkenny, J. D. and Kroll, J. and Landen, O. L. and Liedahl, D. A. and Maclaren, S. A. and Meezan, N. B. and Nikroo, A. and Schneider, M. B. and Thorn, D. B. and Widmann, K. and P\'erez-Callejo, G.},
  journal = {Phys. Rev. Lett.},
  volume = {121},
  issue = {9},
  pages = {095002},
  numpages = {5},
  year = {2018},
  month = {Aug},
  publisher = {American Physical Society},
  doi = {10.1103/PhysRevLett.121.095002},
  url = {https://link.aps.org/doi/10.1103/PhysRevLett.121.095002}
}

@article{JAb96,
year = {1996},
volume = {53},
pages = {705},
author = {J Abdallah and A Ya Faenov and D Hammer and S A Pikuz and G Csanak and R E H Clark},
title = {Electron beam effects on the spectroscopy of satellite lines in aluminum X-pinch experiments},
journal = {Physica Scripta},
doi = {10.1088/0031-8949/53/6/011},
url = {https://doi.org/10.1088/0031-8949/53/6/011},
}

@article{JAb99,
title = {Hot electron effects on the satellite spectrum of laser-produced plasmas},
journal = {Journal of Quantitative Spectroscopy and Radiative Transfer},
volume = {62},
pages = {1-11},
year = {1999},
author = {J. Abdallah and A.Ya Faenov and T.A. Pikuz and M.D. Wilke and G.A. Kyrala and R.E.H. Clark},
doi = {10.1016/S0022-4073(97)00180-5}
}

@article{PhysRevLett.120.055001,
  title = {Origins and Scaling of Hot-Electron Preheat in Ignition-Scale Direct-Drive Inertial Confinement Fusion Experiments},
  author = {Rosenberg, M. J. and Solodov, A. A. and Myatt, J. F. and Seka, W. and Michel, P. and Hohenberger, M. and Short, R. W. and Epstein, R. and Regan, S. P. and Campbell, E. M. and Chapman, T. and Goyon, C. and Ralph, J. E. and Barrios, M. A. and Moody, J. D. and Bates, J. W.},
  journal = {Phys. Rev. Lett.},
  volume = {120},
  issue = {5},
  pages = {055001},
  numpages = {6},
  year = {2018},
  month = {Jan},
  publisher = {American Physical Society},
  doi = {10.1103/PhysRevLett.120.055001},
  url = {https://link.aps.org/doi/10.1103/PhysRevLett.120.055001}
}

\end{document}